\documentclass[aps,floatfix,pre,preprint]{revtex4}
\usepackage{graphicx,bbm,color,bm,changes}
\usepackage{amsmath,amsfonts,amssymb,amsthm}
\usepackage[latin1]{inputenc}

\def\be{\begin{equation}}
\def\ee{\end{equation}}

\newtheorem{thm}{Theorem}
\newtheorem{lm}[thm]{Lemma}
\newtheorem{cl}[thm]{Corollary}

\newtheorem{remark}[thm]{Remark}

\theoremstyle{definition}

\newcommand{\finprf}{\unskip\null\hfill$\square$\vskip 0.3cm}

\newcommand{\rz}{\mathbb{R}}       

\begin{document}
\title{Upper and lower bounds for the speed of fronts of the reaction diffusion equation with Stefan boundary conditions}
\author{Rafael D. Benguria, M. Cristina Depassier}
\affiliation{
Instituto de F\'isica, Pontificia Universidad Cat\'olica de Chile, Av. Vicu\~na Mackenna 4860, Santiago, Chile.
}

\date{\today}

\begin{abstract}
We establish two integral  variational principles for the spreading  speed of the one dimensional  reaction diffusion equation with Stefan boundary conditions. 
The first principle is valid for monostable reaction terms and the second principle is valid  for arbitrary reaction terms.  These principles allow to obtain several  upper and lower bounds for the speed.  In particular, we  construct  a  generalized  Zeldovich-Frank-Kamenetskii type lower  bound  for the speed  and  upper bounds  in terms of the speed of the standard reaction diffusion problem.  We construct 
 asymptotically exact lower bounds previously obtained by perturbation theory.
\end{abstract}

\maketitle

\section{Introduction}

The reaction diffusion equation with a Stefan free boundary condition has received much attention since it was first introduced in \cite{DuLin2010}, where it  was proven that  for the logistic reaction term  a new dynamical effect arises.  In the classical Fisher case any localized  initial  perturbation to the unstable equilibrium state  evolves into a traveling wave of constant speed that spreads to the whole space replacing the initial unstable equilibrium state with a stable state.  When a Stefan  free boundary condition is added it was shown in \cite{DuLin2010} that a vanishing-spreading dichotomy occurs, the initial perturbation may either die out or spread out as in the Fisher problem. It was also shown that for  an initial position of the free boundary  larger than a critical value the initial  perturbation always spreads out,  otherwise, its fate  depends both on the initial perturbation and on the initial position of the free boundary. These results have been extended and shown to be valid for general reaction terms \cite{BuntingDu2012,DuGuo2012,DuLou2015JEMS,DuLou2015Siam,  Sun2015, Hadeler2016, LeiMatsuzawa2018} and for density dependent diffusion coefficients.

 The purpose of this manuscript is to study the speed of the spreading regime for arbitrary reaction terms.   We show that the asymptotic spreading speed can be characterized through  variational principles. This principles allow to obtain as accurate as desired numerical values for the speed and to formulate upper and lower bounds of general validity.

We begin with  a brief summary of some well known results for the classical reaction diffusion problem
\[ u_t = u_{xx} + f(u), \qquad   \mbox{ with }  (x,t) \in  \rz \times \rz_+,  
\]
where the reaction terms $f(u)$ satisfy $f(0)= f(1) = 0$. Depending on additional constraints they are classified  as  monostable, bistable or  of combustion type. 
In the monostable case, which we call case A, it satisfies
  \be\label{caseA}
  f'(0) >0,\,  f'(1) < 0,   \text{ and }   f>0  \text{ on }  (0,1).
  \ee
  It is of bistable type  (case B)  if it satisfies  
 \be
 \label{caseB}
 f(u)  < 0 \text{ for } u \text{ on } (0, a),\,   f  >0 \text{ on } (a, 1),  \text{ and } \int_0^1 f(u) du >0.
 \ee
 In the combustion case (Case C)  it obeys
 \be\label{caseC}
 f=0 \text{ on }  (0,a), \text{ and }  f > 0 \text{ on }  (a, 1).
 \ee
 Finally we include case D, 
  \be\label{caseD}
  f'(0) = 0, \, f>0  \text{ on }  (0,1).
  \ee
  
 The logistic reaction term $f(u) = u(1-u)$  was first studied by Fisher,  and the general  monostable case  studied  by Kolmogorov, Petrovsky and Piskunov \cite{KPP1937} who showed that a sufficient localized initial condition $u(x,0)$ evolves into a monotonic traveling wave $u(x,t) = q(x - c t)$  joining the stable state $u=1$ to the unstable state $u=0$.  The speed was proven to be given by $c_{\textrm{KPP}} = 2 \sqrt{f'(0)}$, for all reaction terms which satisfied $f>0$, $f(u) \le u f'(0 )$  in $(0,1)$ \cite{KPP1937}.  On the other hand,  Zeldovich and Frank-Kamenetskii  (ZFK) \cite{ZFK1938} in the study of combustion studied a reaction term sharply peaked at $u=1$, and for the limiting case of a delta function at $u=1$,  they showed that the speed of the front is given by $c_{\textrm ZFK} = \sqrt{2 \int_0^1 f(u) du}$.  This value was later shown to be a lower bound on the speed for all monostable reaction terms \cite{BerestyckiNirenberg1992} and   later shown to be valid for bistable reaction terms as well.

  Existence of traveling monotonic fronts,  and conditions on initial conditions $u(x,0)$ for  convergence into monotonic fronts were proven in \cite{AronsonWeinberger1978}.
For arbitrary monostable  reaction terms the speed can be determined by  a local \cite{HadelerRothe1975}  or an integral variational principle \cite{BenguriaDepassier1996CMP}. An integral variational principle valid  for  general reaction terms of  any type allows to obtain the speed for cases A-D \cite{BenguriaDepassier1996PRL}.

 In recent years,  the study of a reaction diffusion equation with a free boundary on which Stefan  conditions are applied, has been introduced as a model that exhibits spreading--vanishing dichotomy, effect absent in the standard problem where a sufficiently localized perturbation  $u(x,0)$ always leads to a traveling wave $q(x- c t)$. The dynamics of the scalar reaction diffusion equation with Stefan boundary conditions was first introduced by \cite{DuLin2010,DuLin2013} and further studied in a series of works. See, for example,  \cite{BuntingDu2012,DuGuo2012,DuLou2015JEMS,DuLou2015Siam,  Sun2015, Hadeler2016, LeiMatsuzawa2018}.  
 
 The purpose of this manuscript is to study the asymptotic speed of the traveling wave  in the spreading regime. We construct two integral variational principles  which allow to obtain upper and lower bounds.  In particular we find  explicit upper and lower bounds, valid for all  monostable reaction functions in terms of the ZFK expression, and the
  asymptotic speed $c_0$ of the traveling waves of  the standard reaction diffusion problem ($\kappa \to \infty$),  
namely,

 \be \label{cotas}
\frac{ \kappa} {\kappa + 1}\sqrt{2 \int_0^1 f(u) du} \le c \le \frac{ \kappa} {\kappa + 1} c_0
\ee.

The lower bound is shown to be also valid for bistable reaction functions for which upper bounds are also given.

  \section{Statement of the problem}
  
   We study the reaction diffusion equation in one dimension with Stefan boundary conditions, 
 \begin{align}\label{RDS}
\begin{split}
&u_t = u_{xx} + f(u)   \text{   with  } f(0) = f(1) =0,  u \in (0,1)   \quad x \in (0,L(t)), t>0\\
&\text{with boundary  conditions} \\
&   u_x(0,t) = 0, \quad u(L(t),t) = 0,  \quad  \frac{dL(t)}{dt} = - \kappa u_x(L(t),t), \quad  L(0) = L_0
\end{split}
\end{align}
where $L(t)$ is the free boundary, $\kappa$ is  the Stefan parameter for this problem, and   subscripts denote derivatives with respect to the independent variables $x$ and $t$.  We shall be interested in the spreading regime, that is, the case in which the initial perturbation grows and the front invades the whole space.  It has been shown that in the spreading regime as $t\rightarrow \infty$ the system evolves into a traveling wave solution $u(x,t) = q(x- c t)$ of constant speed, which is unique \cite{DuLou2015JEMS,Sun2015}.
  The problem we address here is the determination of the speed of a spreading front as a function of the reaction term and of the Stefan parameter $\kappa$. 
  Spreading traveling wave solutions $u(x,t) = q ( x - c t)$ correspond to solutions of the ordinary differential equation  \cite{DuLou2015JEMS,Hadeler2016} 
\be\label{eq}
 q_{zz} + c q_z + f(q) = 0, \text{  with   }  q(-\infty) = 1, \, q(0) = 0, \, q_z(0) = -c /\kappa  
 \ee
where $z= x - c t$.

 The approach is based on previous results \cite{BenguriaDepassier1996PRL,BenguriaDepassier1996CMP}
   for the standard reaction diffusion problem.  
   
   Since the asymptotic front $q(z)$ is a decreasing function \cite{DuLou2015JEMS},  we may  work in phase space.  Defining  $p(q) = - q_z  \ge 0 $, solving Eq. (\ref{eq})  reduces to finding the solutions of
   \begin{subequations}
\begin{align}
  &p(q) \frac{d p}{dq}  - c  p(q) + f(q) = 0,  \label{phase}\\
  & p(0) = c/\kappa   \qquad p(1)=0    \qquad p>0 \text{ in } (0,1).   \label{BC}
  \end{align}   
  \end{subequations}
  Notice from (\ref{BC})  that the standard reaction diffusion problem is obtained in the limit $\kappa \to \infty.$

  \section{Variational principle and bounds for monostable reaction terms.}
  
  In this section we prove a variational characterization  for the asymptotic speed of the front. From this characterization we derive upper and lower bounds in terms of the reaction profile.

   \begin{thm}\label{teorema1}
   Let $f \in C^1(0,1)$ with $f(0)=f(1)=0$ and
$f(u)>0$ for $u \in (0,1)$. 
Then,
\begin{equation}\label{VP1}
c = \max_{g\in \mathcal{D}}  \mathcal{E}_{\kappa}(g), 
 \end{equation}
 where
 \begin{equation}
 \label{VP1A}
 \mathcal{E}_{\kappa}(g) = \frac{ 2 \int_0^1 \sqrt{ f(u)  g(u)  h(u)}\, du}{\int_0^1 
 g(u) \, du  + \frac{ g(0)}{\kappa} }.
\end{equation}
In (\ref{VP1})  the maximum is taken over  functions in $\mathcal{D}$, with 
\begin{equation}\label{VP1B}
\mathcal{D} = \{g \in C^1(0,1), g \ge 0, h \equiv - g' >0 \, \mbox{in $ (0,1)$}, g(0) < \infty, g(1)=0\}.
\end{equation}

   \end{thm}
   
\begin{proof}
   
   Let $g(q)$ be an arbitrary decreasing  positive function, so that $h(q ) = - g'(q) >0$. Multiplying  Eq. (\ref{phase}) by $g(q)/p(q)$, and integrating in $q$ between $0$ and $1$, after integrating by parts  and using the boundary conditions  (\ref{BC}) one obtains 
  \be \label{boundmono}
c  \left (\int_0^1 
 g(u) \,du  + \frac{ g(0)}{\kappa}  \right) = \int_0^1 \left(  h(u) p(u) + \frac{f(u) g(u)}{p(u)}\right) \, du \ge 2   \int_0^1 \sqrt{ f(u)  g(u)  h(u)} \,du
\ee
where $ h(u) = -  g'(u) \ge 0.$  The inequality follows from the fact that $h, p, f$ and $g$ are nonnegative and using $a^2 + b^2 \ge 2 a b$. Equation (\ref{boundmono}) implies that $c \ge \mathcal{E}_{\kappa}(g)$ for
 any $g \in \mathcal{D}$.

To conclude the proof of the theorem we need to show that there is a $g \in \mathcal{D}$, say $\tilde g$, such that $c =  \mathcal{E}_{\kappa}(\tilde g)$. For that to happen we must have equality in (\ref{boundmono}). 

Given $p(u)$, the solution of the boundary value problem (\ref{phase}), (\ref{BC}), equality holds in (\ref{boundmono})
for the function $\tilde g$ that satisfies
\begin{equation}\label{equal1}
\tilde h(u) p(u) = \frac{f(u) \tilde g(u) }{p(u) }= \tilde g(u) ( c - p'(u)),
\end{equation}
where we used (\ref{phase}) to get the last equality in Eq. (\ref{equal1}). Since $\tilde h(u) = - \tilde g'(u)$, 
 (\ref{equal1}) can be written as 
 \begin{equation}\label{equal2}
 -\frac{\tilde g'}{\tilde g} = \frac{c}{p} - \frac{p'}{p}.
 \end{equation}
 Integrating  (\ref{equal2}) from $0$ to $u$ we get
 $$
 \log\left( \frac{\tilde g(u)}{\tilde g(0)}\right) = -\int_0^u \frac{c}{ p(s)}\, ds + \log\left( \frac{\tilde p(u)}{\tilde p(0)}\right).
 $$
Using that $p(0) = c/\kappa$ we finally get,
\be \label{tildeg}
\tilde g(u) = \tilde g(0) \,  \frac{ \kappa\,  p(u) }{c}   \exp\left(-  \int_0^u \frac{c}{p(s)} ds\right)
\ee
Moreover, since $p(1)=0,$ it follows from (\ref{tildeg}) that $\tilde g(1) =0.$  We may then  restrict to trial functions that satisfy $g(1) =0$. Due to the homogeneity of $\mathcal{E}_{\kappa}$ in $g$ it follows that $\tilde g(0)>0$ is arbitrary.

It is simple to see from (\ref{tildeg}) that $\tilde g \in \mathcal{D}$. Notice that $\tilde h \ge 0$ follows from the first equality in (\ref{equal1}). 

\end{proof}

 \begin{cl}
 The speed $c$ is a nondecreasing  function of $\kappa$, and $\lim_{\kappa \to 0} c = 0$, $\lim_{\kappa \to \infty} c = c_0$
 \end{cl}
 
   \begin{proof}
   That $c$ is  a nondecreasing function of $\kappa$ follows from the fact that for a fixed $g\in \mathcal{D}$,
   $$
    \mathcal{E}_{\kappa_1} (g) \le  \mathcal{E}_{\kappa_2} (g)
    $$
    if $\kappa_1 < \kappa_2$.
   The limits of the speed follow taking the corresponding limits in (\ref{VP1}).
    \end{proof}
   
   \begin{remark}
  Alternatively, using  the Feynman-Hellman theorem we have, from (\ref{VP1}),
  $$
  \frac{d c}{d \kappa} = \frac{\tilde g(0)}{\kappa^2}  \frac{ 2 \int_0^1 \sqrt{ f(u)  \tilde g(u) \tilde  h(u)}\, du}{\left(\int_0^1 
 \tilde g(u) du  + \frac{ \tilde g(0)}{\kappa} \right)^2} \ge 0.
 $$
  \end{remark}

For any monostable reaction terms we may use trial functions to obtain as accurate as desired lower bounds on the speed. One can also find general upper and lower bounds as we show below.

\subsection{Bounds for monostable reaction functions}

 Here we shall construct the extension of upper and lower bounds established for the speed of the  standard reaction diffusion problem ($\lim_{\kappa \to \infty} c$)  to include the Stefan boundary condition $\kappa >0$. 
  We shall make use of the fact that for any positive decaying function $g(u)$ which adopts a finite value at $u=0$ an such that $g(1) =0$ the following inequalities hold, 
 \be\label{gs}
\frac{\kappa +1}{\kappa} \int_0^1  g(u) du  < \int_0^1 
 g(u) du  + \frac{ g(0)}{\kappa}  < \frac{\kappa +1}{\kappa} g(0).
 \ee
 
 The reaction diffusion equation for combustion problems was studied in \cite{ZFK1938} where, for a reaction term close to a delta function at $u=1$, the speed was found to be  
\be \label{ZFK}
c_{\text{\tiny{ZFK}}} = \sqrt{2 \int_0^1 f(u) du}.
\ee
It was shown later that for all reaction terms of $f\ge 0$ in (0,1) this value represents a lower bound,
\cite{BerestyckiNirenberg1992}, bound which was later proven for more general reaction terms in \cite{BenguriaDepassier1996PRL}.

\begin{lm}
The asymptotic speed of the front is bounded   below by
 \be \label{ZFK1}
c \ge  \frac{ 2 \int_0^1 f(u)  du}{  \int_0^1 du  \sqrt{ 2 \int_u^1 f(s) ds} + \frac{1}{\kappa} \sqrt{ 2 \int_0^1 f(s) ds} } \ge       \frac{ \kappa} {\kappa + 1}\sqrt{2 \int_0^1 f(u) du}
\ee
where $\kappa$ is the Stefan parameter.
\end{lm}

\begin{proof}
A lower bound can obtained by a choice of trial function $g(u)$. 
  Choose the trial function 
  $$
 g(u) = \sqrt{ 2 \int_u^1 f(s) ds}.
$$
It is is easy to verify that for this trial function $ g(u)  h(u) = f(u)$. Using Theorem \ref{teorema1} we obtain 
\be \label{lowerB1}
c \ge \frac
{ 2 \int_0^1 f(u)  du}
{\int_0^1  g(u) du  + \frac{ g(0)}{\kappa} } = 
\frac
{ g^2(0)}
{\int_0^1 
 g(u) du  + \frac{ g(0)}{\kappa} }.
\ee
 Using (\ref{gs}) in (\ref{lowerB1}) we obtain 
\be \label{lowerB1ZFK}
c \ge \frac{\kappa}{\kappa+1} g(0) = \frac{\kappa}{\kappa+1}c_{ZFK}.
\ee
\end{proof}

Notice that the first bound in (\ref{ZFK1}) can be rewritten as
$$
\kappa \le \frac{c  g(0)}{ g^2(0) - c \int_0^1  g(u) du},
$$
which was found as the leading order in an asymptotic expansion for  $c\ll 1$ \cite{Fadai2020}.  Here we have shown that this is not only an asymptotic value for $c\ll 1$ but a lower bound on the speed for all $\kappa$. 
 
\begin{lm}
The asymptotic speed of the front is bounded   above by
 \be \label{upperB1}
c \le \frac{ \kappa} {\kappa + 1} c_0
\ee
where $c_0$ is the speed of the standard reaction diffusion problem with the same reaction function.

\end{lm}

\begin{proof} 
For any positive function $\alpha(u)$ we have that, since $f(u), g(u)$  and $ h(u)$ are positive, 
$
2 \sqrt{f(u) g(u)h(u)} \le \alpha(u) h(u)  + \frac{1}{\alpha(u)} f(u) g(u),
$
Choosing $\alpha(0) =0$ such that $\lim_{u \to 0} f(u)/\alpha(u) =0$ we have
\begin{align*}
2 \int_0^1 \sqrt{ f(u)  g(u)  h(u)} du \le & \int_0^1 \alpha(u)  h(u) du +  \int_0^1 \frac{ f(u)  g(u)}{\alpha(u)}\\
 = & \int_0^1 \left( \alpha'(u) +  \frac{ f(u) }{\alpha(u)}\right)  g(u)  d u \\
  \le  & \sup_{0\le u\le 1} \left( \alpha'(u) +  \frac{ f(u) }{\alpha(u)}\right) \int_0^1 g(u) du.
\end{align*}
Using this inequality in  Theorem \ref{teorema1} we have that,
$$
c\le \max_{ g \in \mathcal{D}}  \frac{ \kappa \sup_u \left[ \alpha'(u) + f(u)/\alpha(u)\right] }
{\kappa + \frac{  g(0)}{\int_0^1  g(u) du}} \le \frac{ \kappa}{\kappa+1} \sup_{0\le u \le 1} \left[ \alpha'(u) + \frac{ f(u)}{\alpha(u)}\right] 
$$
where we used that for $g \in \mathcal{D}$, $g(0) \ge \int_0^1 g(u)\, du$.  Since $\alpha$ is arbitrary,   choosing the function $\alpha$ that gives the least upper bound, we obtain
\be \label{upperHR}
c\le  \frac{ \kappa} {\kappa + 1}\inf_{\alpha} \sup_u \left[ \alpha'(u) + f(u)/\alpha(u) \right],    \text{  where }  \alpha(u) \ge 0 \mbox{ for } u \in (0,1) \mbox{ and } \alpha(0) = 0.
\ee
We identify the right side as the speed of the wave  of the standard reaction diffusion problem (\ref{phase}) with boundary conditions $p(0) = p(1) =0$ \cite{HadelerRothe1975}. 
\end{proof} 

\begin{remark}
Alternatively, using (\ref{gs}) we may write
\be
c = \max_{ g \in \mathcal{D}} \frac{ 2 \int_0^1 \sqrt{ f(u)  g(u)  h(u)} du}{\int_0^1 
 g(u) du  + \frac{ g(0)}{\kappa} } \le \max_{ g \in \mathcal{D}}  \frac{ 2 \int_0^1 \sqrt{ f(u)  g(u)  h(u)} du}{  \frac{\kappa +1}{\kappa}\int_0^1 
 g(u) du } 
\ee
That is,
$$
c\le  \frac{ \kappa} {\kappa + 1} \left[ \max_{ g \in \mathcal{D}}\frac{ 2 \int_0^1 \sqrt{ f(u)  g(u)  h(u)}\, du}{\int_0^1 
 g(u) du }\right].
 $$ 
 We identify the term in the brackets as the speed $c_0$ of the standard reaction diffusion equation \cite{BenguriaDepassier1996CMP}.
 \end{remark}

  \section{Variational principle and bounds for general reaction terms}

  In this section we formulate a variational principle valid for general reaction terms of  types A, B, C and D. 
  
  \begin{thm}
    Let $f \in C^1(0,1)$ with $f(0)=f(1)=0$ and
$f(u)$ belonging to types A, B, C or D. Then
\begin{equation}\label{VP2}
c^2 = \max_{g\in \mathcal{H}}  \mathcal{F}_{\kappa}(g), 
 \end{equation}
 where
 \begin{equation}
 \label{VP2A}
 \mathcal{F}_{\kappa}(g) =  \frac{ 2 \int_0^1 f(u)  g(u)  \,du}{\int_0^1 
\frac{ g^2(u)}{h(u)}\, du  + \frac{ g(0)}{\kappa^2} },
 \end{equation}
 and  the maximum is taken over  functions in $\mathcal{H}$, with
\begin{equation}\label{VP2B}
\mathcal{H} = \{g \in C^1(0,1), g \ge 0, h \equiv - g' >0 \, \mbox{in $ (0,1)$}, \,g(0) < \infty, \, g(1)=0,  \int_0^1 
\frac{ g^2(u)}{h(u)}\, du < \infty \}.
\end{equation}
   
 \end{thm}

\begin{proof}
  
 Let $g(q)$ be an arbitrary decreasing  positive function, so that $h(q ) = - g'(q) >0$. Multiplying  Eq. (\ref{phase}) by $g(q)$, and integrating in $q$ between $0$ and $1$, after integrating by parts and using the boundary conditions  Eq. (\ref{BC}),  one obtains the identity
  \be \label{step1}
  c \int_0^1 p(q) g(q) \,dq -  \frac{1}{2}  \int_0^1  h(q) p^2(q) \,dq = \int_0^1  f(q) g(q)\, dq -   g(0) \, \frac{c^2}{2 \kappa^2}.
  \ee
 Since  $p,\,h,\, f, {\rm and}\, g$ are positive, 
 for every fixed $q$
\be\label{ineq}
c\, p(q) \, g(q) -  \frac{1}{2}    h(q) p^2(q) \le \frac{c^2}{2}  \frac{g^2(q)}{h(q)}
\ee
and replacing (\ref{ineq}) in (\ref{step1}) we obtain
\be\label{bound2}
c^2 \ge \frac{ 2 \int_0^1 f(q) g(q) \,dq}{ g(0)/\kappa^2 + \int_0^1 g^2(q)/h(q)\,  dq} = \mathcal{F}_{\kappa}(g).
\ee
for $g\in \mathcal{H}$. To conclude the proof of the theorem  we must  show that there exists a function $g \in \mathcal{H}$, say $\hat g$, for which equality holds in (\ref{bound2}).
It follows from (\ref{ineq}) that equality holds when $ p(q) = c \hat g(q)/\hat h(q)$, that is, 
\be\label{hatg2}
\frac{\hat g'}{\hat g} = - \frac{c}{p}.
\ee
where $p$ is the solution of (\ref{phase},\ref{BC}).
Since $p(0) > 0$,   the solution of  (\ref{hatg2}) can be written as
\be\label{solhat2}
\hat g(q) = \hat g(0)\,   \text{Exp} \left[- c \int_{0}^q \frac{ds}{p (s)} \right] .
\ee
One can show from (\ref{phase},\ref{BC}) that $p(q)$ behaves as $b (1-q)$ near $q=1$, with $b>0$. It follows then from  (\ref{solhat2}) that $\hat g(1)=0$.  Since (\ref{VP2A}) is homogeneous in $g$, without loss of generality one could choose $ \hat g(0) = 1$, here we choose to leave it as a free parameter. 
 
  \end{proof}

  \begin{cl}
 The speed $c$ is a nondecreasing  function of $\kappa$, and $\lim_{\kappa \to 0} c = 0$, $\lim_{\kappa \to \infty} c = c_0$
 \end{cl}
 
   \begin{proof}
   That $c$ is  a nondecreasing function of $\kappa$ follows from the fact that for a fixed $g\in \mathcal{H}$,
   $$
    \mathcal{F}_{\kappa_1} (g) \le  \mathcal{F}_{\kappa_2} (g)
    $$
    if $\kappa_1 < \kappa_2$.
   The limits of the speed follow taking the corresponding limits in (\ref{VP2}).
    \end{proof}

 \begin{remark}
 Alternatively, using  the Feynman Hellman theorem we obtain
 \be\label{FH}
 \frac{d c^2}{d \kappa} =  \frac{4}{\kappa^3}  \frac{  \int_0^1 f(q) \hat g(q) dq}{ \left[  \int_0^1 \hat g^2(q)/\hat h(q)\,  dq + \frac{1}{\kappa^2}\right]^2} \ge 0
 \ee
 \end{remark}
 Since $c\ge 0$ and $\kappa \ge 0$ it follows that $c$ increases with $\kappa$ for any reaction term. It follows from (\ref{VP2}) that $\lim_{\kappa\to 0} c = 0$ and $\lim_{\kappa\to \infty} c = c_0$, with $c_0$ the speed of the standard reaction diffusion problem with boundary conditions $p(0) = p(1) =0$  for the  same reaction term. 
 
  \finprf
  
\subsection{ Bounds for general reaction functions}

 In this section we show that for the reaction diffusion problem (\ref{RDS}) in cases A, B, C and D,
\be\label{zfkbound}
c\ge \frac{\kappa}{1 +\kappa}  \, c_{\text{\tiny{ZFK}}}.
\ee

To establish this bound we need to choose an appropriate trial function $g$. For the combustion reaction term studied in \cite{ZFK1938} the reaction is close to zero everywhere except close to $u=1$, The solution for $p(u)$ in the region where $f$ is approximately zero is given by 
$p_{\text{\tiny{ZFK}}} (u) =c (1 +  \kappa u)/\kappa $. Using this expression as a guess for $p(u)$ (which is not the solution for arbitrary $f>0$),  we construct a trial function using (\ref{solhat2}).

\begin{lm}
The asymptotic speed of the front   reaction terms of type A, B, C and D is bounded   below by
 \be \label{ZFKgral}
c^2 \ge     \frac{2   \kappa^2}{1+\kappa} \int_0^1 \frac{f(u)}{1 + \kappa u} \,du 
\ge\frac{2   \kappa^2}{(1+\kappa)^2} \int_0^1 f(u) \,du = \left(  \frac{\kappa}{1+\kappa}\, c_{ZFK}\right)^2.
\ee
where $\kappa$ is the Stefan parameter.
\end{lm}

\begin{proof}
Choose the trial function
$$
g = \frac{1}{1 + \kappa u}.
$$
Using this trial function in (\ref{bound2}) we obtain
$$
c^2 \ge \frac{2 \kappa^2}{1+\kappa} \int_0^1 \frac{f(u)}{1 + \kappa u} du 
$$
which  proves the bound (\ref{ZFKgral}).
\end{proof}

In order to obtain upper bounds we will make use of the following inequalities which are proven in the Appendix, 

\be\label{ineq1}
\int_0^1 \frac{g^2(u)}{h(u)} du + \frac{g(0)}{\kappa^2} \ge  \frac{2}{\kappa} \int_0^1  g(u) du \,\,  \text{for} \,\, g(u) >0, \, h(u) = -g'(u) \ge 0, \, g(1)=0,
\ee
and

\be\label{ineq2}
\int_0^1 \frac{g^2(u)}{h(u)} du + \frac{g(0)}{\kappa^2} \ge \frac{4 + \kappa^2}{2 \kappa^2} \int_0^1 u g(u) du,  \,\,  \text{for} \,\, g(u) >0, \, h(u) = -g'(u) \ge 0, \, g(1)=0.
\ee

\medskip
\begin{lm}\label{boundsgrales}
The asymptotic speed of the spreading front for reaction terms of type A, B, C and D satisfies
\be
c\le \min \left( 
\frac{2 \kappa}{\sqrt{4 + \kappa^2}}
\sqrt{\sup_u\left( \frac{f(u)}{u}\right)},
 \,    \sqrt{ \kappa \sup_u f(u)},  \, \sqrt{2 \kappa^2 \sup_u F(u)}
 \right),
\ee
where
$$
F(u) = \int_0^u f(s) ds.
$$ 
\end{lm}

\begin{proof}
First we prove an Aronson-Weinberger type bound, using (\ref{ineq2}).
\be\label{final}
c^2 = \max_g \frac{ 2 \int_0^1 f(q) g(q) dq}{  \int_0^1 g^2(q)/h(q)\,  dq + \frac{g(0)}{\kappa^2}} \le \max_g \frac{4 \kappa^2}{4 + \kappa^2} \frac{ \int_0^1 f(q) g(q) dq}{  \int_0^1  u g(u) du} \le \frac{4 \kappa^2}{4 + \kappa^2} \sup_{0\le u \le 1} \left( \frac{f(u)}{u}\right).
\ee

A similar but less accurate bound can be proved using (\ref{ineq1}). 
\be
c^2 =  \max_g \frac{ 2 \int_0^1 f(q) g(q) dq}{  \int_0^1 g^2(q)/h(q)\,  dq + \frac{g(0)}{\kappa^2}} \le  \kappa  \max_g  \frac{\int_0^1 f(u)  g(u) du}{\int_0^1  g(u) du}  \le \kappa \sup_u f(u).
\ee

A third bound is obtained integrating by parts the numerator, 
\begin{equation*}
\begin{split}
c^2  &= 2 \max_g\frac{\int_0^1 f(u)  g(u) \,du}{  \int_0^1 g^2(q)/h(q)\,  dq + \frac{g(0)}{\kappa^2}}  =  2 \max_g\frac{\int_0^1 F(u) h(u) \, du}{  \int_0^1 g^2(q)/h(q)\,  dq + \frac{g(0)}{\kappa^2}}\\
&\le  \max_g \frac{2 \sup_u F(u) g(0)}{  \int_0^1 g^2(q)/h(q)\,  dq + \frac{g(0)}{\kappa^2}} 
\le 2 \kappa^2 \sup_u F(u).\\
\end{split}
\end{equation*}

\end{proof}

The bound $c \le \sqrt{2 \kappa^2 \sup_u F(u)} $ yields the exact leading asymptotic value of the speed for small $\kappa$ obtained by an asymptotic expansion in (\cite{HachemMcCue2019}) for the reaction term $f(u) = u - u^2$ . For this reaction term $F(u) = u^2/2 - u^3/3$, and $\sup_u F(u) = 1/6$. The bound yields $ c\le \kappa/\sqrt{3}$, which corresponds to the leading order term in the asymptotic expansion   of the velocity for small $\kappa$.

Finally in the next section we apply the variational theory to a specific reaction term which has been studied in the literature in order to obtain a lower bound on the speed.

 \section{Application to the reaction term $\bf{ f(u) = u^m(1-u)} $}
 
For the  reaction term $f(u) = u^m(1-u)$, $m>0$,  it has been shown \cite{Sun2015} that the the speed of the Stefan front is continuous and strictly decreasing in $m$. For $m=1$ the speed as a function of $\kappa$ is studied numerically in \cite{HachemMcCue2019}  and an explicit analytical expression is found by means of perturbation theory  close to $c=0$ \cite{HachemMcCue2019}. Here we prove the first result using the variational principle (\ref{VP2}) and obtain an analytical lower bound for the speed for the case $m=1$.

That the speed is a decreasing function of $m$ follows from the variational principle (\ref{VP2}) and the fact that for $0<u<1$, $u^m$ is decreasing in $m$ therefore, for a fixed $g$,
$$
\int_0^1 u^m (1-u) g(u) \, du$$
is a decreasing function of $m$ therefore $c$ is also a decreasing function of $m$.
 
Next we apply the results given above  to the Fisher equation $f(u)  = u (1 - u)$. 
We use two trial functions to obtain lower bounds on the speed. First consider the simplest decaying function that satisfies $g(0) =1$, namely
$$
g_1(q) = 1- \lambda q,
$$
which for $\lambda= 0.22$ yields the bound
\be \label{simple}
c\ge c_1 = \frac{0.23141 \kappa}{1 + 0.0556737 \kappa^2}, 
\ee
To construct a second trial function we use  a guess   $p(q)  =  (c/\kappa)(1-q)(1 + \kappa q)$ and use (\ref{solhat2}) we construct the second trial function
$$g_2(q) =  \left( \frac{1-q}{1+ \kappa q}\right)^{\kappa/(1+\kappa)}
$$  
With this trial function (\ref{VP2}) yields 
\be \label{better}
 c^2 \ge  \frac{2 \kappa^2 (1 + \kappa) 
  \left[
   (3 + 4 \kappa) \left(  2 -\,  {_2}F_1[1, \kappa_1, 3 + \kappa_1,-\kappa ]  \right) - (2+\kappa)\, \, {_2}F_1[1, \kappa_1, 4 + \kappa_1,-\kappa ] \right]  }
 { 
 (1+2 \kappa)(3 + 4\kappa ) \left( 2 + 3 \kappa + \kappa (1 + \kappa)\, \, {_2}F_1[ 1,\kappa_2, 3 + \kappa_1,-\kappa ] \right)  }
  \ee
where $ \kappa_1 = \kappa/(\kappa+1), \kappa_2 =   -1/(\kappa+1). $

 In Fig. 1 we compare these bounds   with the numerical results obtained in \cite{HachemMcCue2019}. 
 
  \begin{figure}[h!] \label{FigHachem}
\centering
\includegraphics[width=0.45 \textwidth]{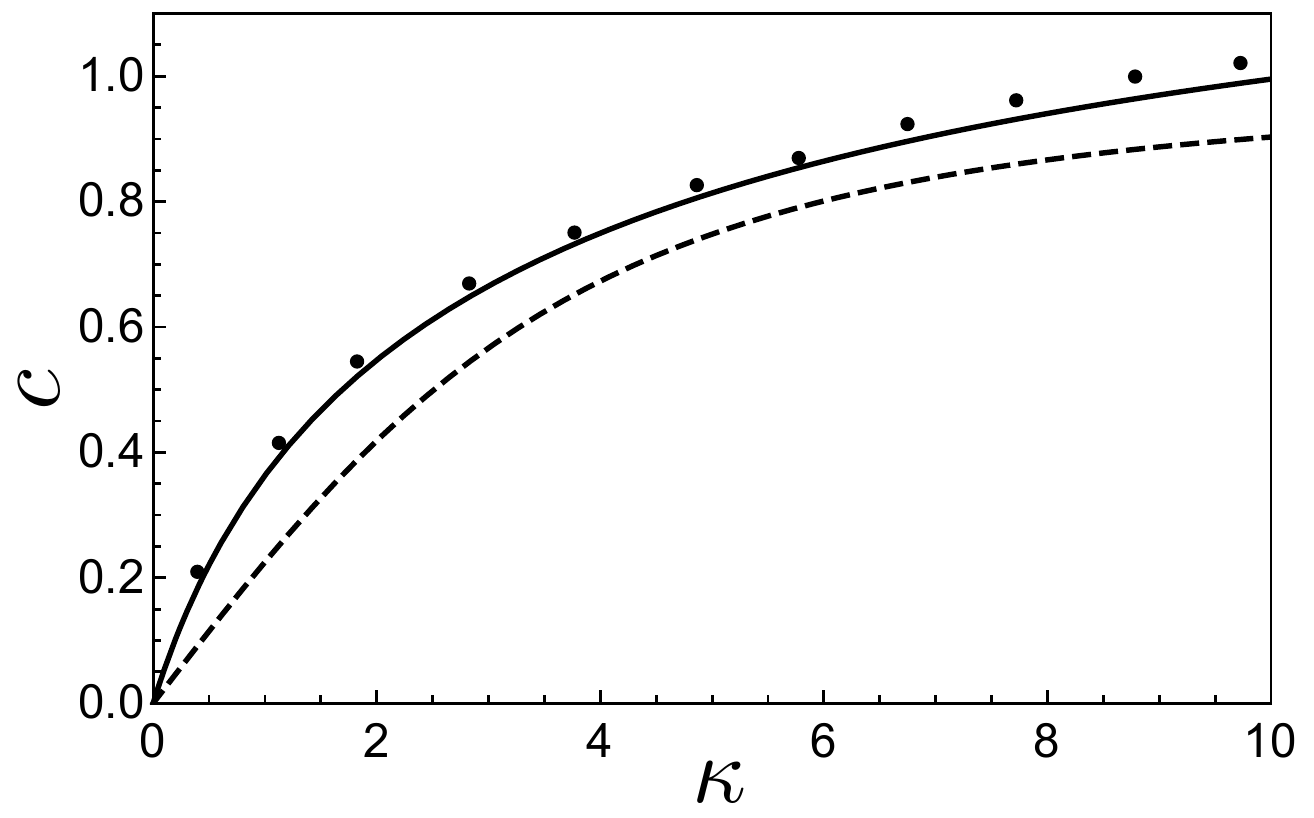}
\caption{
Speed as a function of $\kappa$ for the Fisher reaction term. The continuous line is the lower bound  (\ref{better}) and the dashed lined is the bound  (\ref{simple}) obtained with the simplest trial function.   The dots show the numerical results reported in \cite{HachemMcCue2019}.
 }
\end{figure}
We see that the lower bound is close to the numerical results, the exact value can be approached arbitrarily close by adequate choice of trial functions.

\section{Remarks}

The variational principles that we have formulated  can be extended straightforwardly to include a nonlinear convective term   of the form $\phi(u) u_x$ and to the case of density dependent reaction diffusion  with Stefan boundary conditions. 

\section{Acknowledgments}
This work has been partially supported by Fondecyt project (Chile) 1201055.

\section{Appendix}

In this Appendix we prove inequalities used above.

Proof of (\ref{ineq1}).

 \begin{align}
 D \equiv \int_0^1 \frac{g^2(u)}{h(u)}\, du + \frac{g(0)}{\kappa^2} &= \int_0^1 \frac{g^2(u)}{h(u)} + \frac{h(u)}{\kappa^2} \,  du  \\
 &\ge 2  \int_0^1 \sqrt{\frac{ g^2(u)}{\kappa^2}}\, du =  \frac{2}{\kappa} \int_0^1 g(u) \, du.
 \end{align}
 To get the second equality we used the definition of $h$ and the fact that $g(1)=0$. To obtain the inequality we used that $h>0$.
 
Proof of (\ref{ineq2}).

First notice that 
\be\label{App4}
g(0) \ge 2 \int_0^1 u g(u) du
\ee
which follows from $g(0)>g(u)$.  In addition 
\be\label{App5}
\int_0^1 \frac{ g^2(u)}{h(u)}  du \ge \frac{1}{2} \int_0^1  u  g(u) du,
\ee
To prove this write
\begin{align*}
\left(  \int_0^1 u g(u) \,du \right)^2 &= \left( \int_0^1 \frac{g(u)}{\sqrt{h(u)}} (u \sqrt{h(u)} )\, du\right)^2  \\
                  &\le  \int_0^1 \left( \frac{g(u)}{\sqrt{h(u)}}\right)^2  \, du \int_0^1 \left(u \sqrt{h(u)}\right)^2 \, du \\
                  & \le  2 \int_0^1 \frac{g^2(u)}{h(u)} \, du \int_0^1 u g(u)\, du,
\end{align*}
where in the last term we integrated by parts and used the fact that $g(1)=0$ and $g(0)$ is finite.      Using 
(\ref{App4}) and (\ref{App5}) we obtain  the inequality (\ref{ineq2}).

 \end{document}